\documentclass{jnmp}

\begin{document}
\setcounter{page}{311}
\renewcommand{\evenhead}{S~Yu~Sakovich}
\renewcommand{\oddhead}{Addendum to: ``Coupled KdV Equations \dots''}

\thispagestyle{empty}


\FirstPageHead{8}{2}{2001}
{\pageref{sakovich:add-firstpage}--\pageref{sakovich:add-lastpage}}{Addendum}

\copyrightnote{2001}{S~Yu~Sakovich}

\Name{Addendum to:  ``Coupled KdV Equations\\
 of Hirota-Satsuma Type''
(J.~Nonlin. Math. Phys. Vol. 6, No.3 (1999), 255--262)}\label{sakovich:add-firstpage}

\Author{S~Yu~SAKOVICH}

\Address{Department of Physics, Middle East Technical University, 06531 Ankara, Turkey\\[2mm]
Permanent address:\\
Institute of Physics, National Academy of Sciences, 220072 Minsk, Belarus\\
E-mail: saks@pisem.net}

\Date{Received February 12, 2001; Accepted  March 10, 2001}

\begin{abstract}
\noindent
It is shown that one system of coupled KdV equations, found in
{\it J.~Nonlin.~Math.~Phys.}, 1999, V.6, N~3, 255--262 to possess the Painlev\'{e}
property, is integrable but not new.
\end{abstract}

In our recent paper \cite{sakovich-add:Sak}, we found that the system of coupled KdV
equations
\begin{equation}\arraycolsep=0em
\begin{array}{l}
u_{t}=u_{xxx}+9v_{xxx}-12uu_{x}-18vu_{x}-18uv_{x}+108vv_{x},
\vspace{1mm}\\
v_{t}=u_{xxx}-11v_{xxx}-12uu_{x}+12vu_{x}+42uv_{x}+18vv_{x}
\end{array}
\label{sakovich-add:v}
\end{equation}
passes the Painlev\'{e} test for integrability well, but we were unable to
find a parametric zero-curvature representation for this system there. In this
addendum, we show that the system (\ref{sakovich-add:v}) \emph{is integrable but not new}:
it is related by a simple transformation of variables to an integrable system
introduced a long time ago by Drinfeld and Sokolov~\cite{sakovich-add:DS}.

In their paper, in Example~13, Drinfeld and Sokolov gave the Lax
representation $L_{t}=\left[  A,L\right]  $ with the differential operators
\begin{equation}\arraycolsep=0em
\begin{array}{l}
L=\left(  D^{3}+2uD+u_{x}\right)  \left(  D^{2}+v\right)  ,
\vspace{2mm}\\
\displaystyle A=D^{3}+\left(  \frac{6}{5}u+\frac{3}{5}v\right)  D+\left(  -\frac{3}{5}
u_{x}+\frac{6}{5}v_{x}\right)
\end{array}
\label{sakovich-add:la}
\end{equation}
for the system of coupled KdV equations
\begin{equation}\arraycolsep=0em
\begin{array}{l}
\displaystyle u_{t}=-\frac{4}{5}u_{xxx}+\frac{3}{5}v_{xxx}-\frac{12}{5}uu_{x}+\frac{3}
{5}vu_{x}+\frac{6}{5}uv_{x},\vspace{2mm}\\
\displaystyle v_{t}=\frac{3}{5}u_{xxx}-\frac{1}{5}v_{xxx}+\frac{12}{5}vu_{x}+\frac{6}
{5}uv_{x}-\frac{6}{5}vv_{x}.
\end{array}
\label{sakovich-add:ds}
\end{equation}

It is easy to see that the transformation
\begin{equation}
t\rightarrow10t,\qquad u\rightarrow-\frac{3}{2}u+\frac{3}{2}v,\qquad
v\rightarrow-2u-3v
\label{sakovich-add:tr}
\end{equation}
changes the system (\ref{sakovich-add:ds}) into the system (\ref{sakovich-add:v}).
This solves the
problem of integrability of the system (\ref{sakovich-add:v}). Moreover, using the scalar
spectral problem $L\phi=\lambda\phi$, $\phi_{t}=A\phi$, where the operators $L
$ and $A$ are given by (\ref{sakovich-add:la}) and $\lambda$ is a parameter, and the
transformation (\ref{sakovich-add:tr}), we can construct the first-order linear problem
$\Psi_{x}=X\Psi$, $\Psi_{t}=T\Psi$, or the zero-curvature representation
$X_{t}=T_{x}-\left[  X,T\right]  $, for the system (\ref{sakovich-add:v}), with the
following $5\times5$ matrices $X$ and $T$:
\[
X=\left(
\begin{array}{ccccc}
0 & 1 & 0 & 0 & 0\\
2u+3v & 0 & 1 & 0 & 0\\
0 & 0 & 0 & 1 & 0\\
0 & 0 & \frac{3}{2}\left(  u-v\right)  & 0 & 1\\
\lambda & 0 & 0 & \frac{3}{2}\left(  u-v\right)  & 0
\end{array}
\right)  ,
\]
$T=\{\{5u_{x}-15v_{x},-10u+30v,0,10,0\},\,\{5u_{xx}-15v_{xx}-20u^{2}
+30uv+90v^{2},-5u_{x}+15v_{x},5u+15v,0,10\},\,\{10\lambda,0,30v_{x}
,-30v,0\},\,\{0,10\lambda,30v_{xx}-45uv+45v^{2},0,-30v\},\,\{5\lambda
u+15\lambda v,0,10\lambda,30v_{xx}-45uv+45v^{2},-30v_{x}\}\}$, where the
cumbersome matrix $T$ is written by rows.

We can conclude now, that \emph{all} the systems of coupled KdV equations,
which passed the Painlev\'{e} test in \cite{sakovich-add:Sak}, have turned out to be
integrable.

\medskip

The author is deeply grateful to Prof.~Atalay Karasu for a copy of \cite{sakovich-add:DS}
and numerous useful discussions, and to the Scientific and Technical Research
Council of Turkey (T\"{U}B\.{I}TAK) for support.

\label{sakovich:add-lastpage}
\end{document}